\documentclass[useAMS,usenatbib]{mn2e}
\usepackage{graphicx}
\usepackage{times}

\title[Ice chemistry in massive Young Stellar Objects: the role of metallicity]
{Ice chemistry in massive Young Stellar Objects: the role of metallicity}
\author[Oliveira et al.]{J.M.\,Oliveira$^{1}$\thanks{joana@astro.keele.ac.uk}, 
J.Th.\,van Loon$^{1}$, G.C.\,Sloan$^{2}$, R.\,Indebetouw$^{3,4}$, 
F. Kemper$^{5}$, \and A.G.G.M.\,Tielens$^{6}$, J.D.\,Simon$^{7}$, 
Paul\,M.\,Woods$^{5}$, M.\,Meixner$^{8}$\\
$^{1}$ Lennard-Jones Laboratories, School of Physical and Geographical Sciences,
Keele University, Staffordshire ST5 5BG, UK\\
$^{2}$ Department of Astronomy, Cornell University, Ithaca, NY 14853, USA\\
$^{3}$ Department of Astronomy, University of Virginia, P.O.\ Box 400325, 
Charlottesville, VA 22904, USA\\
$^{4}$ National Radio Astronomy Observatory, 520 Edgemont Road, Charlottesville,
VA 22903, USA\\
$^{5}$ Jodrell Bank Centre for Astrophysics, School of Physics and Astronomy, 
The University of Manchester, Manchester M13 9PL, UK\\
$^{6}$ Leiden Observatory, P.O.\ Box 9513, NL-2300 RA Leiden, The Netherlands\\
$^{7}$ Observatories of the Carnegie Institution of Washington, 813 Santa 
Barbara St., Pasadena, CA 91101, USA\\
$^{8}$ Space Telescope Science Institute, 3700 San Martin Drive, Baltimore, MD 
21218, USA\\}

\begin{document}

\date{Accepted 2010 November 10. Received 2010 November 10; in original form 
2010 September 7}

\pagerange{\pageref{firstpage}--\pageref{lastpage}} \pubyear{2010}

\maketitle

\label{firstpage}

\begin{abstract} We present the comparison of the three most important ice
constituents (water, CO and CO$_2$) in the envelopes of massive Young Stellar 
Objects (YSOs), in environments of different metallicities: the Galaxy, the 
Large Magellanic Cloud (LMC) and, for the first time, the Small Magellanic Cloud
(SMC). We present observations of water, CO and CO$_2$ ice in 4 SMC and 3 LMC 
YSOs (obtained with Spitzer-IRS and VLT/ISAAC). While water and CO$_2$ ice are 
detected in all Magellanic YSOs, CO ice is not detected in the SMC objects. 
Both CO and CO$_2$ ice abundances are enhanced in the LMC when compared to 
high-luminosity Galactic YSOs. Based on the fact that both species appear to be
enhanced in a consistent way, this effect is unlikely to be the result of 
enhanced CO$_2$ production in hotter YSO envelopes as previously thought. 
Instead we propose that this results from a reduced water column density in the 
envelopes of LMC YSOs, a direct consequence of both the stronger UV radiation 
field and the reduced dust-to-gas ratio at lower metallicity. 
In the SMC the environmental conditions are harsher, and we observe a reduction 
in CO$_2$ column density. Furthermore, the low gas-phase CO density and higher 
dust temperature in YSO envelopes in the SMC seem to inhibit CO freeze-out.
The scenario we propose can be tested with further observations.
\end{abstract} 


\section{Introduction}

The most abundant astronomical ices are water, CO and CO$_2$. These ices involve
abundant metals that, as gas-phase atoms and molecules, are prime coolants of 
the molecular clouds \citep[e.g.][]{omukai10}. Efficient cooling during the 
early collapse stages is paramount to allowing star formation to proceed. By 
changing the chemical balance in the molecular cloud, interstellar ice chemistry
is thus an important ingredient in the formation of stars from contracting 
cores. 

In spite of its ubiquity and significance as a cornerstone of interstellar 
chemistry, the formation of water is poorly understood. In comparison to 
gas-phase water abundance, the observed water ice abundance is too high to have 
formed by direct accretion from the gas-phase only \citep{roberts02}. 
Therefore surface reactions on cold dust grains likely play a significant role 
in the formation of water molecules \citep{tielens82}. 

Water ice seems to form very easily in molecular clouds, as it requires 
relatively low gas densities and only moderately low temperatures 
\citep[e.g.][]{hollenbach09}. CO, the most abundant gas-phase molecule besides 
molecular hydrogen, forms exclusively in the gas phase. It is thought that a 
layer of water ice (with impurities of CO and other molecules) forms on cold 
dust grains. During pre-stellar collapse the increased density causes virtually 
all gas species to accrete onto the icy grains, creating a layer dominated 
mainly by CO \citep[e.g.][]{aikawa01}. Thus the CO ice profiles towards 
quiescent regions are dominated by pure CO or CO-rich (apolar) matrices, with
an increasing contribution of CO in water-rich (or polar) matrices towards YSO 
environments \cite[e.g.][]{gibb04}. CO$_2$ forms exclusively via surface 
reactions \citep{millar91} and the observed profiles suggest that these 
reactions occur both in the CO- and water-dominated layers, the latter being the
dominant component \citep[$\ga 50$\%,][]{gerakines99} towards all environments. 

In this letter we investigate the effect of metallicity on the abundance of the
three most important contributors to interstellar ices: water, CO and CO$_2$.
Our research is based on observations in the Magellanic Clouds (MCs) which have
subsolar metallicity \citep[$Z_{\rm LMC} \sim \frac{1}{2} Z_{\odot}$ and 
$Z_{\rm SMC} \sim \frac{1}{5} Z_{\odot}$, e.g.][]{russell89}. We present the 
first observations of CO and CO$_2$ ice towards Young Stellar Objects (YSOs) in 
the SMC. We compare the ice abundances towards luminous YSOs in the Galaxy, LMC 
and SMC, and relate these to the observed properties of the Interstellar Medium 
(ISM), e.g. gas-phase abundances, dust temperature and radiation field strength.
We propose a simple schematic scenario that can explain the observed relative 
ice column densities and the non-detection of CO ice in the SMC environments.

\section{Spectroscopic observations}

\begin{figure}
\centering{
\includegraphics[scale=0.52]{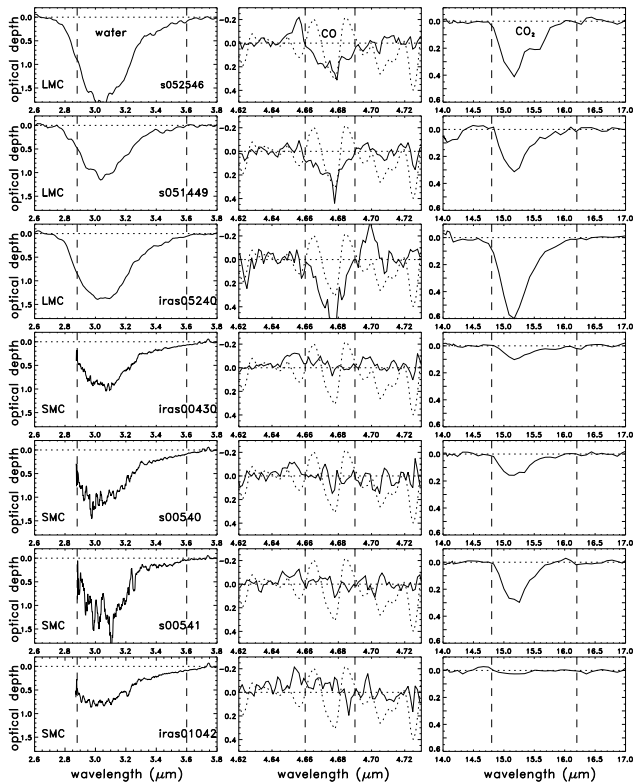}}
\caption{Optical depth spectra of the Magellanic YSOs. The standard star 
spectrum (dotted line) shows the telluric line at $\sim4.68$\,$\mu$m that 
affects the CO ice feature. Dashed lines show the column density integration 
intervals. No CO ice is detected towards the SMC. The water spectra of the SMC
objects have been smoothed in this figure.}
\label{spectra}
\end{figure}

The sample of 7 Magellanic YSOs (Table\,\ref{table1}) includes 3 LMC objects 
described by \citet{oliveira09} and \citet{shimonishi10} and 4 SMC sources. 
IRAS\,01042$-$7215 in the SMC was identified as a YSO by \citet{vanloon08}. The 
three other SMC sources are spectroscopically identified as YSOs for the first 
time in this work, and are amongst the brightest of a sample of YSO candidates 
in the SMC (Oliveira et al. in preparation), identified at 70\,$\mu$m in the 
S$^3$MC survey \citep{bolatto07}. The spectral energy distributions of these 
YSOs suggest that they are young and embedded, and of similar total luminosity 
(a few $10^4$\,L$_{\odot}$). In terms of luminosity, the MC sample is comparable
to the Galactic sample of massive YSOs observed with ISO \citep[]
[and references therein]{gibb04}.

M-band spectra were obtained with the Infrared Spectrometer And Array Camera
(ISAAC) at the European Southern Observatory (ESO)'s Very Large Telescope (VLT) 
on the nights of 4 and 5 November 2009 (ESO Programme 084.C-0955, P.I. J.M.
Oliveira), to search for the CO ice feature at $\lambda=4.67$ $\mu$m. The 
low-resolution mode of ISAAC was used with a $1^{\prime\prime}$-wide slit, 
resulting in a spectral resolution of $\Delta\lambda\approx0.012$ $\mu$m around 
the CO feature, sampled by 0.0014 $\mu$m pixels. Exposure times varied between 
45 and 90 min. The standard IR technique of chopping and nodding was employed 
to cancel the high background. Spectra of standard stars were obtained and
were used to cancel telluric absorption in the YSO spectra; telluric lines were 
used to calibrate the wavelength scale to an accuracy 
$\Delta\lambda\approx0.003$ $\mu$m.

The L-band spectra of three of the SMC sources were obtained with ISAAC at the
ESO/VLT on the nights of 28 and 29 October 2006 (ESO Programme 078.C-0338,
P.I. J.M. Oliveira), the spectrum of IRAS\,01042$-$7215 was published by 
\citet{vanloon08}. The L-band spectra were obtained and reduced in the same way 
as the M-band spectra. The resolving power is $\lambda/\Delta\lambda\approx500$.
Exposure times varied between 60\,$-$\,105 min. The hydrogen lines in the 
standard stars left remnants of at most a few per cent of the continuum level. 
Telluric lines were used to calibrate the wavelength scale to an accuracy 
$\Delta\lambda\approx0.002$ $\mu$m. The 3$-$4\,$\mu$m spectra of the LMC sources
were obtained with the AKARI satellite \citep[published by][]{shimonishi10}. 

All sources were observed with the InfraRed Spectrograph (IRS) on board the 
Spitzer Space Telescope, either as part of the SAGE-Spec Spitzer Legacy Program 
of the LMC \citep{kemper10} and already described in \citet{oliveira09}, or as 
part of GTO programme \#50240 on the SMC (P.I. G.C. Sloan) and shown here for 
the first time. Details on the data reduction of the IRS spectra are found in
\citet{sloan08}.

Continuum subtraction was performed by fitting a $2^{\rm nd}$- to $5^{\rm
th}$-degree polynomial over a narrow wavelength range surrounding the ice
features. Fig.\ \ref{spectra} shows the optical depth spectra. We detect 
water, CO and CO$_2$ ices towards all 3 LMC sources; \citet{shimonishi10} did
not detect CO ice towards IRAS\,05240$-$6809. We detect water and CO$_2$ towards
the SMC sources (only at a 3-$\sigma$ level for IRAS\,01042$-$7215); the most 
striking observation is the absence of CO ice in any of the SMC spectra 
(Fig.\,\ref{spectra} middle). 


\section{Results: ice column densities}

\begin{table*}
\caption{Target positions and ice column densities (in $10^{17}$ molecule 
cm$^{-2}$).}
\label{table1} 
\small{     
\centering          
\begin{tabular}{@{\hspace{2mm}}l@{\hspace{2mm}}lllll}     
\hline\hline   
\hspace{-2mm}Galaxy&Target ID&\multicolumn{1}{c}{RA \& Dec (J2000)}&\multicolumn{1}{c}{N(H$_2$O)}&\multicolumn{1}{c}{N(CO)}&\multicolumn{1}{c}{N(CO$_2$)}\\
\hline
LMC&SSTISAGE1C\,J051449.41$-$671221.5 &05:14:49.41\,$-$67:12:21.5& 19.3 $\pm$ 1.0 &  1.60 $\pm$ 0.11 & 6.10 $\pm$ 0.45\\
LMC&IRAS\,05240$-$6809                &05:23:51.14\,$-$68:07:12.4& 27.5 $\pm$ 1.5 &  2.62 $\pm$ 0.22 &\llap{1}2.92 $\pm$ 0.26\\
LMC&SSTISAGE1C\,J052546.49$-$661411.3 &05:25:46.52\,$-$66:14:11.3& 33.6 $\pm$ 1.4 &  1.54 $\pm$ 0.16 & 8.93 $\pm$ 0.18\\
SMC&IRAS\,00430$-$7326 &00:44:56.30\,$-$73:10:11.8& 17.7 $\pm$ 0.7 &\multicolumn{1}{c}{$\le$ 0.26}& 2.08 $\pm$ 0.09\\
SMC&S3MC\,00540$-$7321 &00:54:02.31\,$-$73:21:18.6& 21.6 $\pm$ 0.8 &\multicolumn{1}{c}{$\le$ 0.19}& 3.50 $\pm$ 0.19\\
SMC&S3MC\,00541$-$7319 &00:54:03.36\,$-$73:19:38.4& 22.3 $\pm$ 1.2 &\multicolumn{1}{c}{$\le$ 0.16}& 5.40 $\pm$ 0.16\\
SMC&IRAS\,01042$-$7215                &01:05:49.29\,$-$71:59:48.8& 16.6 $\pm$ 0.7 &\multicolumn{1}{c}{$\le$ 0.45}& 0.49 $\pm$ 0.18\\
\hline                  
\end{tabular}}
\end{table*}

\begin{figure*}
\includegraphics[scale=0.32]{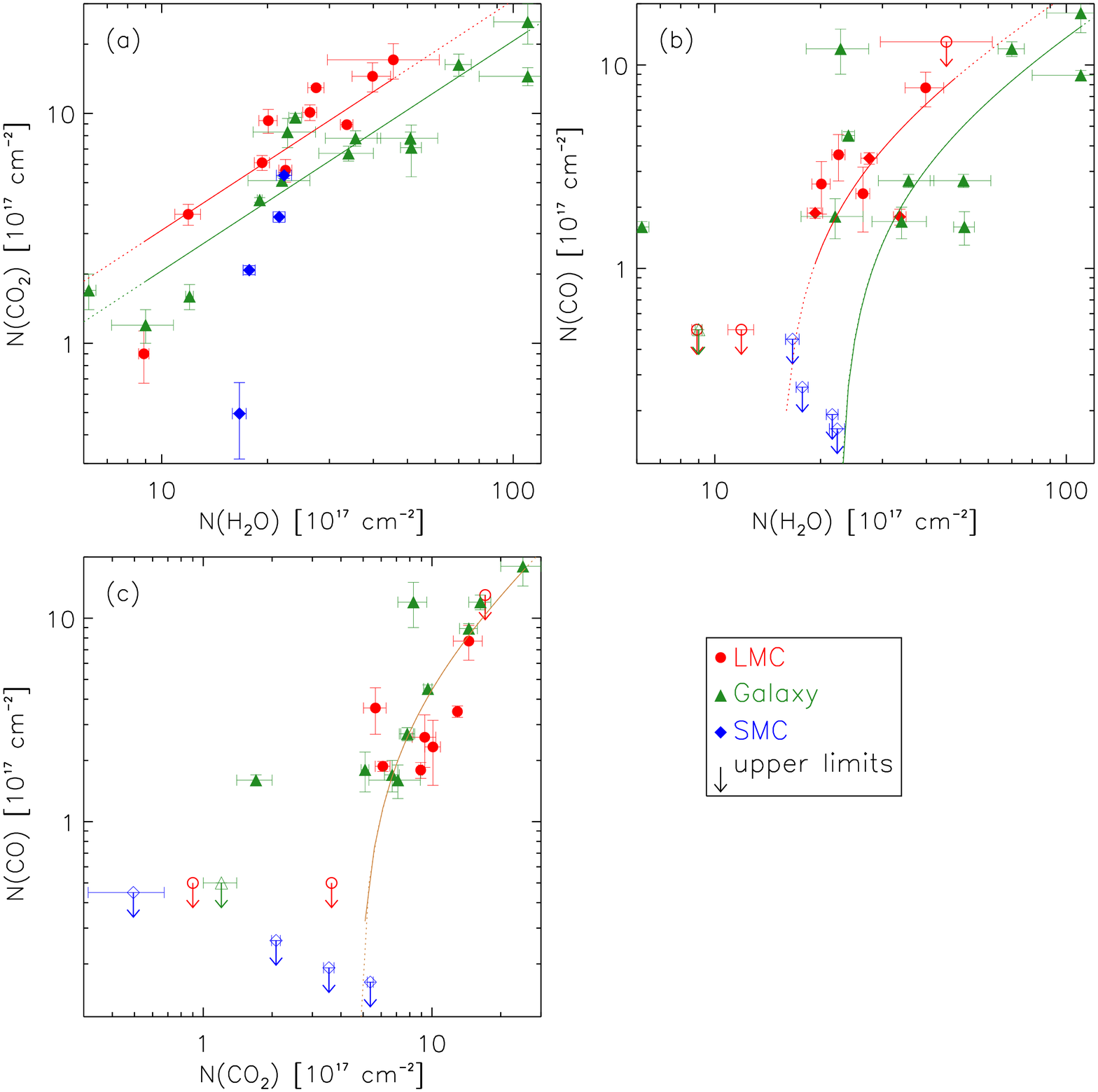}
\hspace{-23mm}
\includegraphics[scale=0.65]{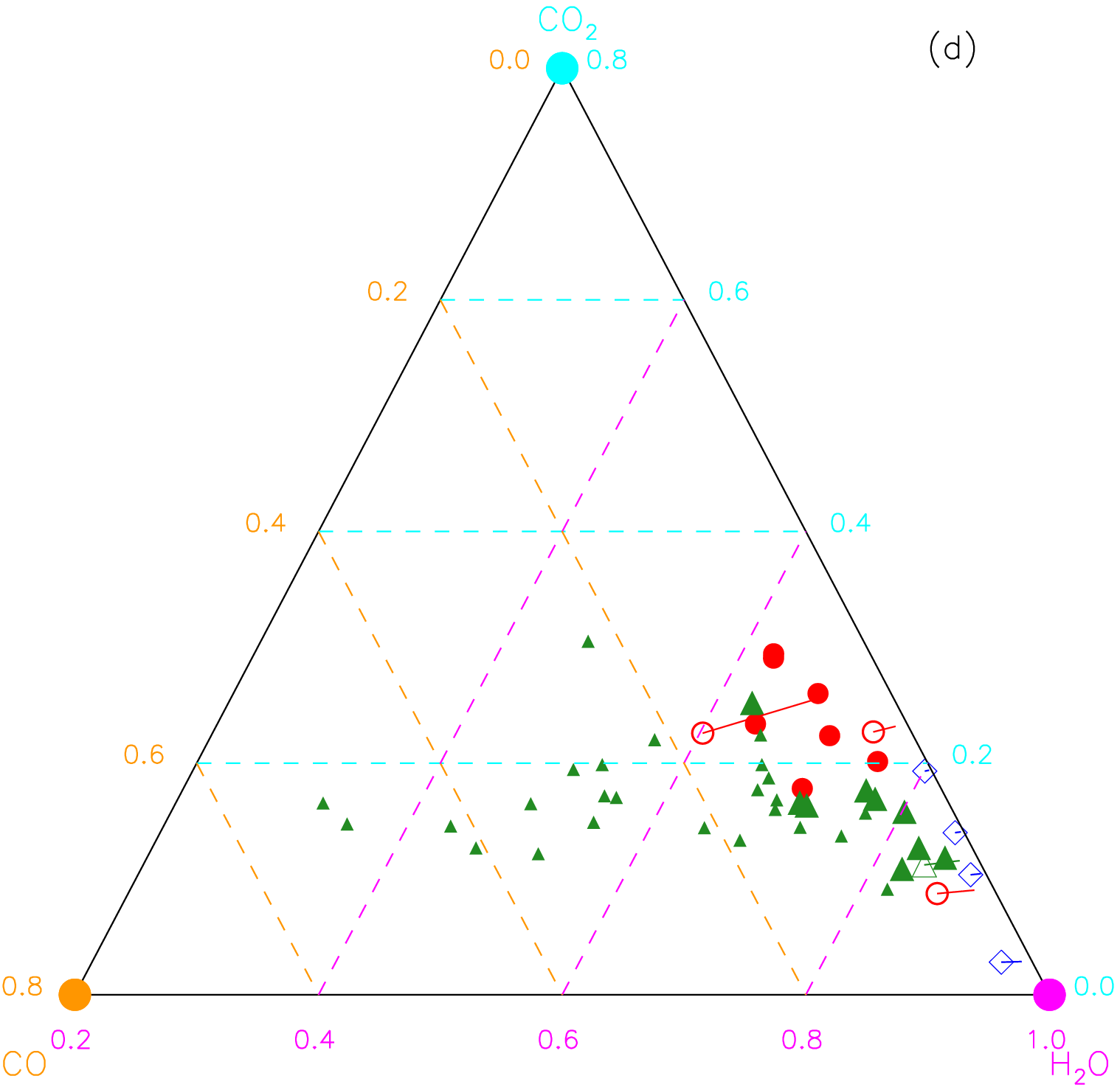}
\caption{Ice column densities. Galactic data (triangles) are from 
\citet{gerakines99}, \citet{gibb04} and \citet[][low-luminosity YSOs, smaller 
triangles]{pontoppidan08}. LMC data (circles) are from this work and 
\citet{shimonishi10}; SMC data (diamonds) are from this work. The solid lines in
panels a, b and c are line fits in linear space. The ternary plot for the column
densities is also shown (panel d, see text); in this diagram the lines
associated with the CO non-detections (open symbols) indicate the change in 
object position if the CO column density is a third of its upper limit.}
\label{cds}
\end{figure*}

We measured column densities for the three ice species by integrating the
optical depth spectra over fixed wavelength ranges (2.9$-$3.6, 4.66$-$4.69 and 
14.8$-$16.2\,$\mu$m). Fig.\,\ref{spectra} shows that we miss the
extreme red wing of the H$_2$O ice profile, which accounts for $\sim$10\% of the
total column density. The integration interval for the CO ice excludes the 
Pf$\beta$ hydrogen recombination line at $\lambda=4.65$ $\mu$m and the S(9) line
from molecular hydrogen at $\lambda=4.687$ $\mu$m. Adopted band strengths are
$2.0\times10^{-16}$, $1.1\times10^{-17}$ and $1.1\times10^{-17}$ cm per
molecule, respectively for H$_2$O, CO and CO$_2$ \citep{gerakines95}. For the 
LMC objects, we compare our measurements to those by \citet{shimonishi10}. Our 
water and CO$_2$ column densities are generally smaller (by 10$-$20\%), due to 
different continuum and integration interval choices. For CO ice, our column
densities are lower by as much as 50\%. This is likely due to the lower 
resolution of the AKARI spectra (R\,$\sim$\,80, compared to R\,$\sim$\,390 for
our spectra) that makes continuum determination uncertain. Such effects 
are not accounted for in the statistical errors.

Fig.\,\ref{cds} shows the ice column densities for the MC samples (from this
work and \citealt{shimonishi10}) and the massive YSO Galactic samples 
\citep{gerakines99,gibb04} for the 3 ice species analysed here --- we only 
consider objects with measurements or upper limits for all 3 species. The 
N(CO$_2$)/N(H$_2$O) ratios for the LMC and the Galaxy are consistent with 
previous determinations \citep{gerakines99,oliveira09,shimonishi10}, 
respectively $\sim$\,0.32 and $\sim$\,0.2 (Fig.\,\ref{cds} (a)). Thus CO$_2$ ice
seems to be more abundant (with respect to water ice) in the LMC than in the 
Galaxy. For the SMC sources, this ratio is lower than those of the other 
samples: in comparison to both LMC and Galactic objects with similar water 
column density (16.6\,$-$22.3\,$\times 10^{17}$ molecules cm$^{-2}$), the SMC 
sources exhibit smaller CO$_2$ ice column density. The sample size and water 
column density range sampled are however small. Nevertheless, this suggests that
while CO$_2$ ice abundance is enhanced in the LMC, it is depleted in the SMC.

CO ice is detected for most of the LMC sources. In Fig.\,\ref{cds} (c) CO and 
CO$_2$ column densities are plotted against each other. The LMC and Galactic 
sources are indistinguishable in this diagram, and there is a strong correlation
(Pearson coefficient $\sim$\,0.94). With the exception of a single source, no CO
ice is detected for sources with CO$_2$ column density $\la 5$\,$\times 10^{17}$
molecule cm$^{-2}$. This is expected since CO is the most volatile of the ice 
species considered here. The slope of the linear relationship is $\sim$\,0.84, 
and a CO$_2$ column density threshold for the detection of CO of 
$\sim$\,4.7\,$\times 10^{17}$ molecule cm$^{-2}$ is inferred \citep[similar to 
those derived by][]{gerakines99}. Since for a given CO$_2$ ice abundance, 
Galactic and LMC objects show similar CO ice abundances, this means that both 
species are affected in a similar way by whatever causes the CO$_2$ enhancement 
with respect to water ice.

CO and water column densities are clearly correlated for the LMC and Galactic 
samples (Fig.\,\ref{cds} (b), Pearson correlation coefficient of $\sim$\,0.8), 
but there is considerable scatter. Even though less clear than in the case of
CO$_2$, LMC sources tend to exhibit larger CO column densities than Galactic 
sources for the same water column density. Due to the larger scatter, we do not 
fit these data; instead we derive the linear relations that are consistent with 
those observed between CO and CO$_2$, and CO$_2$ and water. This results in 
slopes of 0.27 and 0.17 respectively for the LMC and the Galaxy, and water 
column density thresholds for the presence of CO of $\sim$ 14.5 and 23\,$\times 
10^{17}$ molecule cm$^{-2}$.

Since water, CO and CO$_2$ are the most abundant ice species \citep[e.g.][]
{vandishoeck04}, the relationship between the 3 ice column densities can be
visualised using a ternary plot \citep{cook10}. This is shown in Fig.\,\ref{cds}
(d), where column densities are shown as a percentage of the total ice column 
density. Each vertex of the triangle represents mixtures of pure water ice
(magenta), 80\% CO ice (orange) and 80\% CO$_2$ (cyan), with the abundance of a 
particular species decreasing with increasing distance to the vertex. For each 
YSO, the relative ice abundance of each species can be read by following the 
dashed lines and reading the scale with the corresponding colour (see above). 

The LMC and Galactic YSOs exhibit different ice properties as they clearly 
separate in the ternary diagram. As suggested above, CO and CO$_2$ vary in a 
concerted way. Therefore the only way for the LMC sample to be made to occupy 
the same region in this diagram as the Galactic sample is for either the CO and 
CO$_2$ abundances to be reduced, or the water column density to be increased
(roughly by a factor 1.6$-$2). In other words in the LMC either CO and CO$_2$ 
ices are overabundant or water ice is depleted. In Fig.\,\ref{cds} (d) we also 
show the ice abundances of low-luminosity Galactic YSOs 
\citep[small green circles,][]{pontoppidan03,pontoppidan08}. Low- and
high-luminosity (massive) YSOs separate well in this diagram: massive YSOs are 
located further away from the CO vertex, an effect of CO desorption 
\citep{cook10} related to the higher temperatures reached in their envelopes. 
Furthermore, the Galactic low-luminosity and LMC objects do not occupy the 
same region in the ternary diagram. This cautions against the similarities in 
CO$_2$ ice properties found in these samples by \citet{oliveira09}, and 
reinforces the importance of jointly analysing the 3 major ice species. To avoid
the added complication of evolutionary effects in the YSO envelopes, our 
analysis is based on the comparison of the MC sample to massive Galactic YSOs, 
that are in the same evolutionary stages \citep{oliveira09,shimonishi10}.

None of the SMC sources show CO ice in their spectrum (upper limits are provided
in Table\,\ref{table1} and are shown in Fig.\,\ref{cds}). This seems to suggest 
that CO ice does not form or is swiftly desorbed around SMC YSOs. However, only
for one SMC YSO is the CO$_2$ column density above the CO$_2$ threshold 
($\sim$\,4.7\,$\times 10^{17}$ molecule cm$^{-2}$) for the detection of CO ice 
(Table\,\ref{table1}). We believe CO ice is depleted in the SMC (the upper 
limits are stringiest in the sources with largest water and CO$_2$ column
densities) but our data are not yet conclusive. In the ternary diagram the SMC 
sources obviously appear at the limit of low CO column densities.

\vspace*{-3mm}


\section{Discussion}

Of the three ice species investigated here, CO is the most volatile: desorption
temperatures are $\sim$\,20$-$30, $\sim$\,50$-$90, and $\sim$\,100$-$120\,K, 
respectively for CO, CO$_2$, and H$_2$O, depending on the ice mixtures in which
the species reside \citep{fraser01,vanbroekhuizen06}. The largest contribution 
to CO$_2$ ice column density arises from chemical reactions within the 
water-rich ice layer, both in Galactic and LMC objects 
\citep{gerakines99,oliveira09}. On the other hand, the largest contribution to 
CO column density comes from the CO-rich layer that freezes out on top of the 
water-rich layer, at higher densities deeper within the cloud \citep[][at least 
for Galactic YSOs]{gibb04}. Thus, CO survives only in the more shielded regions 
\citep{hollenbach09}. Consequently, environmental changes in the molecular 
clouds can change the balance of species in the icy layers.

The ternary diagram (Fig.\,\ref{cds} (d)) clearly shows that ice abundances
towards YSOs in the LMC are distinct from those towards Galactic YSOs. To 
explain the larger N(CO$_2$)/N(H$_2$O) ratio observed in the LMC, both 
\citet{oliveira09} and \citet[][see also \citealt{shimonishi08}]{shimonishi10} 
suggest that CO$_2$ production is enhanced in the LMC. These authors propose 
this is due to the stronger UV field \citep{welty06} and higher dust 
temperatures \citep{bernard08} observed in the LMC, that have been shown in 
laboratory experiments to increase CO$_2$ production 
\citep{dhendecourt86,ruffle01}. While CO$_2$ forms in the icy grain mantles, CO 
forms in the gas-phase and freezes out onto cold dust grains. If the temperature
of the grain surfaces is on average higher as proposed above, CO ice abundance 
would be decreased. Thus it is difficult to see how solid CO abundance can be 
enhanced in the harsher ambient conditions in the LMC.

\begin{figure}
\centering
\includegraphics[scale=0.32]{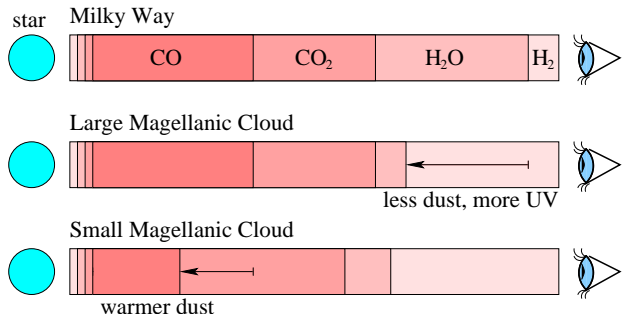}
\caption{Schematic of the column densities (not to scale) of the three ice 
species (for a given H$_2$ column) compared between the Galaxy, LMC, and SMC.
Within the molecular cloud, the outer region (light pink) of the YSO envelope
contains only water ice (the least volatile species) while the shielded interior
region (dark pink) contains all three ice species. The main observed effects are
lower water ice column going from the Galaxy to the LMC, and a reduced CO ice 
column when moving on to the SMC.}
\label{diagram}
\end{figure}

If both CO$_2$ and CO ice abundances appear enhanced towards LMC objects but in
such a way that relative CO-to-CO$_2$ abundances are unchanged (see 
Fig.\,\ref{cds} (c)), another possibility is that the water ice column density 
is reduced towards YSOs in the LMC. Indeed, the largest water column density 
shown in Fig.\,\ref{cds} are $\sim$\,50 and $\sim$\,100\,$\times 10^{17}$ 
molecule\,cm$^{-2}$, respectively in the LMC and the Galaxy (lower in the SMC), 
even though this could be a selection effect. We explore a scenario 
(Fig.\,\ref{diagram}) to explain the reduction of the water ice column density 
at lower metallicity.

Water ice forms easily in molecular clouds. However, accumulation of water ice 
only occurs once there is enough shielding and the rate of ice formation 
overcomes the rate of photo-desorption (assuming the dust grains are cold enough
and thermal desorption is negligible). The A$_{\rm V}$-threshold for water ice 
freeze-out (A$_{{\rm V}f})$ is affected by the strength of the incident 
radiation field \citep{hollenbach09}, varying between 
$A_{{\rm V}f}$\,$\sim$\,3\,mag (for 
typical Galactic molecular clouds) to $A_{{\rm V}f}$\,$\sim$\,6\,mag. The 
reduced metallicity of the MCs compared to the Galaxy results in a lower 
dust:gas ratio (e.g. LMC, \citealt{bernard08}; SMC, \citealt{leroy07}); this 
combined with the harsher UV interstellar radiation field \citep[e.g.][]
{welty06} probably leads to enhanced photo-desorption of water and reduced 
freeze-out in the outer H$_2$ envelopes of molecular clouds in the MCs. Hence 
the ``snowline'' for water moves inwards in the YSO envelope, reducing the 
column density of water ice as observed in the MCs (Fig.\,\ref{diagram}). A 
reduced availability of water-rich ice is also in agreement with the lower 
fraction of CO$_2$ locked in the polar component as found by \citet{oliveira09}.

Within the shielded denser regions of the YSO envelopes, CO$_2$ and CO are 
abundant in the LMC, and we see no significant difference between Galactic and
LMC YSOs in terms of these ices. CO ice is even more sensitive to shielding than
water ice (the A$_{\rm V}$-threshold for the detection of CO ice in Galactic
sources is$> 6$\,mag, \citealt{bergin05,whittet07}). However, gas-phase CO also 
requires high density to accrete onto the icy grains. For the same amount of 
dust (i.e. equivalent shielding) the corresponding gas density is higher in the 
LMC \citep[c.f.][]{vanloon10b} due to the lower dust:gas ratio. These two 
effects could balance in such a way that CO freeze-out still occurs. 

In the Galaxy, the A$_{\rm V}$-threshold for the detection of CO$_2$ ice is 
$\sim 4$\,mag \citep{bergin05,whittet07}, consistent with that of water ice 
within the large uncertainties. Our scenario relies on an extinction threshold 
difference between the two species, that is not currently detected in the 
Galaxy. However, it is possible that such a difference is more pronounced in 
the low metallicity/high UV environments prevalent in the MCs. Even if CO$_2$ 
ice is also affected by the reduced shielding in the outer cold regions, the 
higher density argument put forward in the previous paragraph would also favour 
an enhancement in CO-based CO$_2$ formation in the LMC, as it is indeed observed
\citep{oliveira09}. 

In the SMC the CO$_2$ ice abundance seems depressed and no CO ice is detected 
(see caveats discussed in the previous section). This could be a direct effect 
of the lower CO gas-phase abundance \citep[e.g.][]{leroy07}, and thus a clear 
metallicity effect. Furthermore, \citet{vanloon10b} found that typical dust 
temperatures within YSO envelopes in the SMC are higher than in the LMC. Both 
these effects seem to prevent significant freeze-out of gas-phase CO in the SMC.
Hence YSOs in the SMC show reduced CO$_2$ and CO ice column densities when 
compared to water ice. 

\vspace*{-5mm}
\section{Summary}

We find that the observed column densities of CO$_2$, CO and water ice towards 
LMC YSOs are more consistent with the depletion of water ice, not an increased 
CO$_2$ production as previously proposed. This depletion can result from the
combined effects of a lower gas-to-dust ratio and stronger UV radiation field,
that forces the onset of water ice freeze-out deeper into the YSO envelope,
reducing the observed column density. This trend may continue on to the SMC. 

We present the first CO and CO$_2$ ice spectra of massive YSOs in the SMC. 
CO$_2$ ice seems to be depleted and no CO ice is detected, possibly due to the 
lower gas-phase CO abundance, and higher dust temperatures in the SMC YSOs.
\vspace*{-5mm}
\section*{Acknowledgments}
We thank the staff at ESO-Paranal for their support during the observations. We 
thank T. Shimonishi for the AKARI spectra and A.D. Bolatto for comments. This 
work is based on observations made with the Spitzer Space Telescope, which is 
operated by the Jet Propulsion Laboratory, California Institute of Technology 
under contract with NASA.

\bsp

\label{lastpage}

\end{document}